\documentclass[aip,jcp,amsmath,reprint,superscriptaddress,a4paper, floatfix]{revtex4-1}
\usepackage[english]{babel}
\usepackage{graphicx}
\usepackage{amssymb}
\usepackage{amsmath}
\usepackage{amsfonts}
\usepackage{color}

\usepackage[update,prepend]{epstopdf}

\def\figwidth{\columnwidth}

\newcommand{\la}{\left\langle}
\newcommand{\ra}{\right\rangle}
\begin{document}

\title{Confinement-induced fractionation and liquid-liquid phase separation of polymer mixtures}
\author{Arash Nikoubashman}
\email{anikouba@uni-mainz.de}
\affiliation{Institute of Physics, Johannes Gutenberg University Mainz, Staudingerweg 7, 55128 Mainz, Germany}
\affiliation{Department of Mechanical Engineering, Keio University, Yokohama 223-8522, Japan}

\author{Miho Yanagisawa}
\affiliation{Komaba Institute for Science, Graduate School of Arts and Sciences and Center for Complex Systems Biology, Universal Biology Institute, The University of Tokyo, Meguro, Tokyo 153-8902, Japan}
\affiliation{Graduate School of Science, The University of Tokyo, Bunkyo, Tokyo 113-0033, Japan}

\date{\today}

\begin{abstract}
The formation of (bio)molecular condensates {\it via} liquid-liquid phase separation in cells has received increasing attention, as these coacervates play important functional and regulatory roles within biological systems. However, the majority of studies focused on the behavior of pure systems in bulk solutions, thus neglecting confinement effects and the interplay between the numerous molecules present in cells. To advance our knowledge, we perform simulations of binary polymer mixtures in droplets, considering both monodisperse and polydisperse molecular weight distributions for the longer polymer species. We find that confinement induces a spatial separation of the polymers by length, with the shorter ones moving to the droplet surface. This partitioning causes a distinct increase of the local polymer concentration in the droplet center, which is more pronounced in polydisperse systems. Consequently, the systems exhibit liquid-liquid phase separation at average polymer concentrations where bulk systems are still in the one-phase regime.
\end{abstract}

\maketitle

The liquid-Liquid phase separation (LLPS) of (bio)molecules in living cells has attracted much attention as a mechanism for intracellular organization {\it via} formation of biomolecular condensates.\cite{brangwynne:sci:2009, hyman:biorev:2014} To elucidate the underlying mechanisms of LLPS, bulk solutions of purified biomolecules from cells have been analyzed extensively over the past decade. Although such {\it in vitro} studies facilitate the analysis of LLPS and comparison with simple theoretical models,\cite{martin:sci:2020} they often ignore the molecular diversity encountered in cellular environments.\cite{zimmermann:bio:1991} For example, broad molecular weight distributions can have a profound impact on the phase behavior of polymers, {\it e.g.}, leading to the self-assembly of monodisperse micelles from polydisperse surfactants\cite{mantha:prm:2019} or the fractionation of polymer chains by molar mass in solutions near criticality.\cite{heukelum:mm:2003, zhao:jc:2016}

The demixing of polymers typically becomes even more pronounced in confinemenet due to the associated spatial inhomogeneity.\cite{nikoubashman:jcp:2021} For example, simulations of binary polymer mixtures in spherical droplets have revealed an entropy-driven spatial segregation of the confined polymers based on their molecular weight, stiffness, and/or topology.\cite{zhou:poly:2019, zhou:poly:2020, howard:jcp:2020} Experimentally, confinement effects have been studied using cell-sized water-in-oil droplets containing polymer mixtures, such as polyethylene glycol (PEG) and bovine serum albumin, PEG and DNA, or PEG and Dextran.\cite{yanagisawa:lng:2022, yanagisawa:bpr:2022} Recently, Watanabe {\it et al.} studied the phase coexistence of PEG-Dextran mixtures in cell-sized droplets, discovering that the two-phase coexistence region in small droplets extends to much lower PEG and Dextran concentrations compared to bulk systems.\cite{watanabe:acsml:2022} The bulk behavior was only recovered for rather large droplets with radii $R > 20\,\mu{\rm m}$. They speculated that this $R$-dependent phase separation stemmed from a confinement-induced partitioning of the polymers. To elucidate the origin of this confinement-induced LLPS, we simulate PEG-Dextran mixtures in spherical confinement at two different droplet radii.

To faithfully reproduce experimental conditions, we first need to determine the interaction parameters for the PEG (P), Dextran (D), and water (W) particles (see Methods section). Following previous simulation studies,\cite{groot:bpj:2001, luo:jcr:2012} we use a Flory-Huggins interaction parameter of $\chi_{\rm P-W} = 0.3$ for the PEG-water interactions, which has been extracted from experimental phase coexistence measurements by Saeki {\it et al.}\cite{saeki:poly:1976} We take $\chi_{\rm D-W} = 0.50$ for the Dextran-water interactions, derived from experimental vapor-pressure measurements at $T=298\,{\rm K}$ conducted by Bercea {\it et al.}\cite{bercea:macp:2011} This value is in excellent agreement with previous findings by Clark,\cite{clark:cp:2000} who applied a Flory-Huggins theory-based analysis to experimental tie line data of PEG-Dextran mixtures in water at $T=298-300\,{\rm K}$;\cite{edmond:bcj:1968, king:aiche:1988} Clark also extracted the PEG-Dextran interaction parameter from his analysis, {\it i.e.}, $\chi_{\rm P-D} = 0.031 \pm 0.007$. Note that the physically relevant quantity for a pair of polymers is the {\it combined} Flory-Huggins parameter $\chi N$,\cite{rubinstein:book:2003} and thus a scaled interaction parameter $\chi_{\rm P-D}^{\rm eff} \approx 1.23$ needs to be used in the simulations to reach the same $(\chi N)_{\rm P-D} \approx 100$ as in the experiments. Given the uncertainties in extracting $\chi_{ij}$ from experiments, and the high degree of coarse-graining of our model, it is, however, unclear whether this description reproduces faithfully the interactions between the PEG and Dextran chains. 

\begin{figure}[htb]
    \centering
    \includegraphics[width=\figwidth]{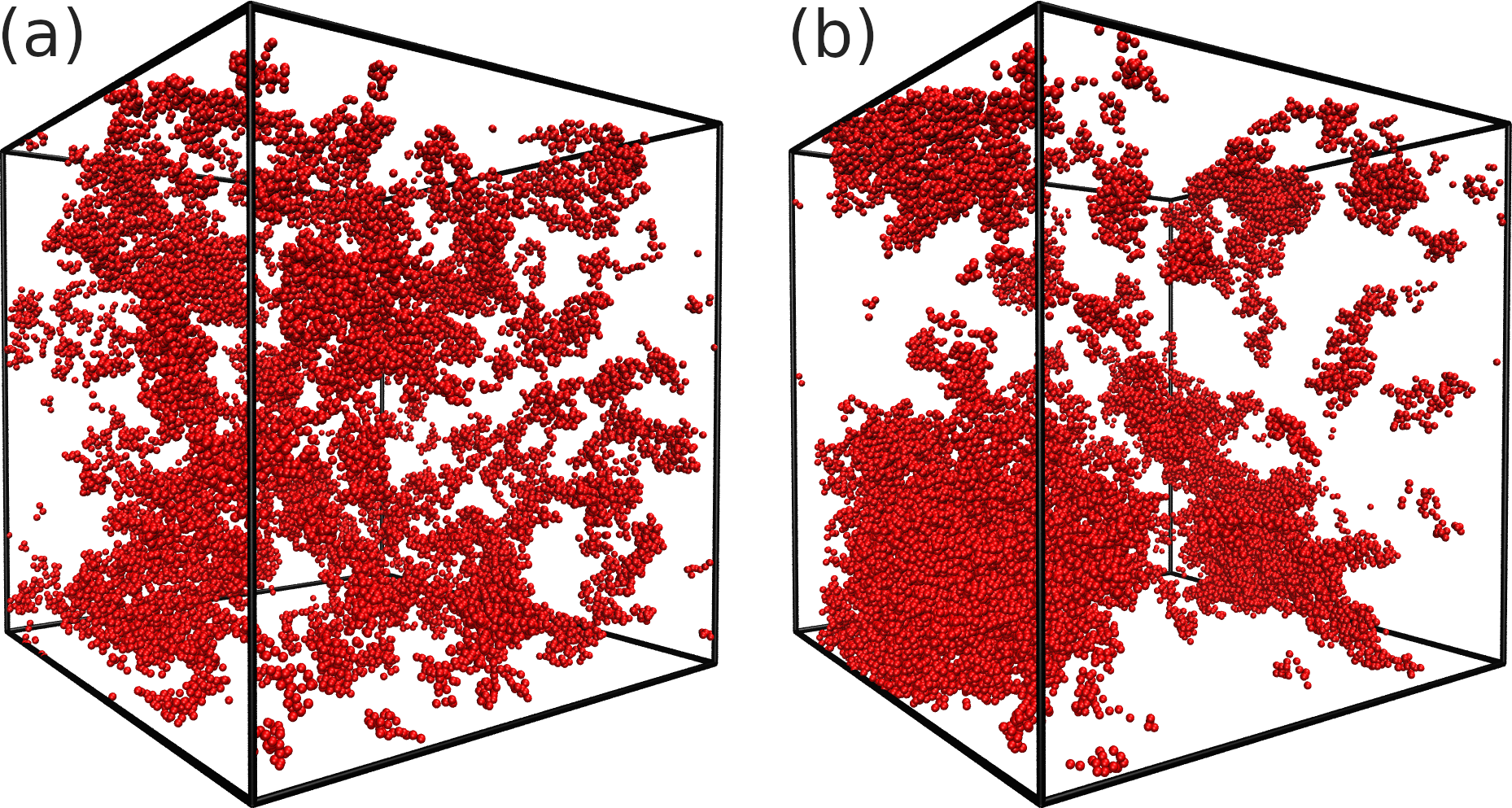}
    \caption{Simulation snapshot of the bulk system at (a) $c_{\rm P} = 3\,{\rm wt}\%$ and $c_{\rm D} = 3\,{\rm wt}\%$, and (b) $c_{\rm P} = 4\,{\rm wt}\%$ and $c_{\rm D} = 4\,{\rm wt}\%$. Only Dextran beads shown for clarity. Snapshots rendered using Visual Molecular Dynamics (version 1.9.3).\cite{vmd}}
    \label{fig:snapBulk}
\end{figure}

To test (and tune, if necessary) the parameterization of our coarse-grained model, we first attempt to reproduce the experimentally known phase behavior\cite{liu:lng:2012, watanabe:acsml:2022} of aqueous PEG-Dextran mixtures in the bulk. The groups of Dimova\cite{liu:lng:2012} and Yanagisawa\cite{watanabe:acsml:2022} determined the binodals of mixtures containing short PEG chains ($M_{\rm w}^{\rm P} = 6$ or $8\,{\rm kg/mol}$) and long Dextran chains ($M_{\rm w}^{\rm D} \approx 500\,{\rm kg/mol}$), finding a critical concentration of slightly below $4\,{\rm wt}\%$ PEG and $4\,{\rm wt}\%$ Dextran. We simulate bulk systems at two concentrations, {\it i.e.}, $c_{\rm P} = 3\,{\rm wt}\%$ PEG and $ c_{\rm D} = 3\,{\rm wt}\%$ Dextran, where we expect a single phase, and at $c_{\rm P} = 4\,{\rm wt}\%$ and $c_{\rm D} = 4\,{\rm wt}\%$, where the mixtures should phase separate. We perform simulations in a cubic box with an edge length of $L \approx 360\,{\rm nm}$ and apply periodic boundary conditions to all three Cartesian directions. The systems are initialized by placing all PEG particles and Dextran chains in opposite halves of the simulation box, and are then run until the density profiles do not change anymore. For $\chi_{\rm P-D}^{\rm eff} = 1.23$, we find a single phase at both concentrations, which indicates that the chosen $\chi_{\rm P-D}^{\rm eff}$ value is too small. Therefore, we iteratively increase $\chi_{\rm P-D}^{\rm eff}$ until we observe aggregation of the Dextran chains at the higher concentration, and a fully mixed system at the lower concentration. This is achieved for $\chi_{\rm P-D}^{\rm eff} = 18.6$ (see simulation snapshots in Fig.~\ref{fig:snapBulk}), which is about 15 times larger than our initial estimate for the PEG-Dextran interactions. The discrepancy between the initial estimate and the final value of $\chi_{\rm P-D}^{\rm eff}$ is rather large, and we can only speculate about its origin: Clark\cite{clark:cp:2000} extracted $\chi_{\rm P-D}$ from experimental coexistence curves using Flory-Huggins solution theory, which ignores the polymer architecture, and thus the branching of the Dextran chains. Further, a monodisperse molecular weight distribution is assumed in his treatment, although generally available Dextran polymers typically have a broad molecular weight distribution.\cite{li:jacs:2008, zhao:jc:2016, liu:fic:2019, yanagisawa:bpr:2022, watanabe:acsml:2022} Finally, we use a rather coarse-grained description, which maps about 40 monomers onto a single bead, resulting in a larger entropy of mixing in the simulations compared to the experiments.\cite{rubinstein:book:2003} To compensate these effects, we will use $\chi_{\rm P-D}^{\rm eff} = 18.6$ in the following.

In recent experiments by Watanabe {\it et al.}, confinement-induced phase separation of PEG-Dextran mixtures was observed for droplets with radii $R < 20\,\mu{\rm m}$.\cite{watanabe:acsml:2022} Simulating such large droplets is computationally infeasible, even at the employed level of coarse-graining (see Methods section), as roughly $7 \times 10^9$ particles would already be needed to represent a droplet with $R=5\,\mu{\rm m}$. Therefore, we perform simulations at two smaller radii, {\it i.e.}, $R \approx 260\,{\rm nm}$ and $R \approx 380\,{\rm nm}$, which should still allow us to capture the effect of confinement on the phase behavior. Further, we consider mixtures containing either monodisperse or polydisperse Dextran chains, since high molecular weight Dextran usually has a broad molecular weight distribution; for example, the Dimova group used Dextran chains with $M_{\rm w} \approx 380-490\,{\rm kg/mol}$ and dispersities in the range of $\DJ \equiv M_{\rm w}/M_{\rm n} \approx 1.8-2.2$,\cite{li:jacs:2008, liu:lng:2012, zhao:jc:2016} while the Yanagisawa group used Dextran with $M_{\rm w} \approx 500\,{\rm kg/mol}$ and $\DJ \approx 3.1$.\cite{watanabe:acsml:2022, yanagisawa:bpr:2022} In our simulations with polydisperse Dextran chains, we draw the molecular weight of each polymer from a Gaussian distribution, targeting $\DJ \approx 1.5$ and $M_{\rm n} \approx 500\,{\rm kg/mol}$. The PEG chains are kept monodisperse throughout, which is consistent with the rather small polydispersity of $\DJ \approx 1.1$ reported in the experimental literature for low molecular weight PEG.\cite{li:jacs:2008, zhao:jc:2016, yanagisawa:bpr:2022} In all simulations, we pick the number of PEG and Dextran chains so that $c_{\rm P} = 3\,{\rm wt}\%$ and $c_{\rm D} = 3\,{\rm wt}\%$, averaged over the entire droplet volume.

\begin{figure}[htb]
    \centering
    \includegraphics[width=\figwidth]{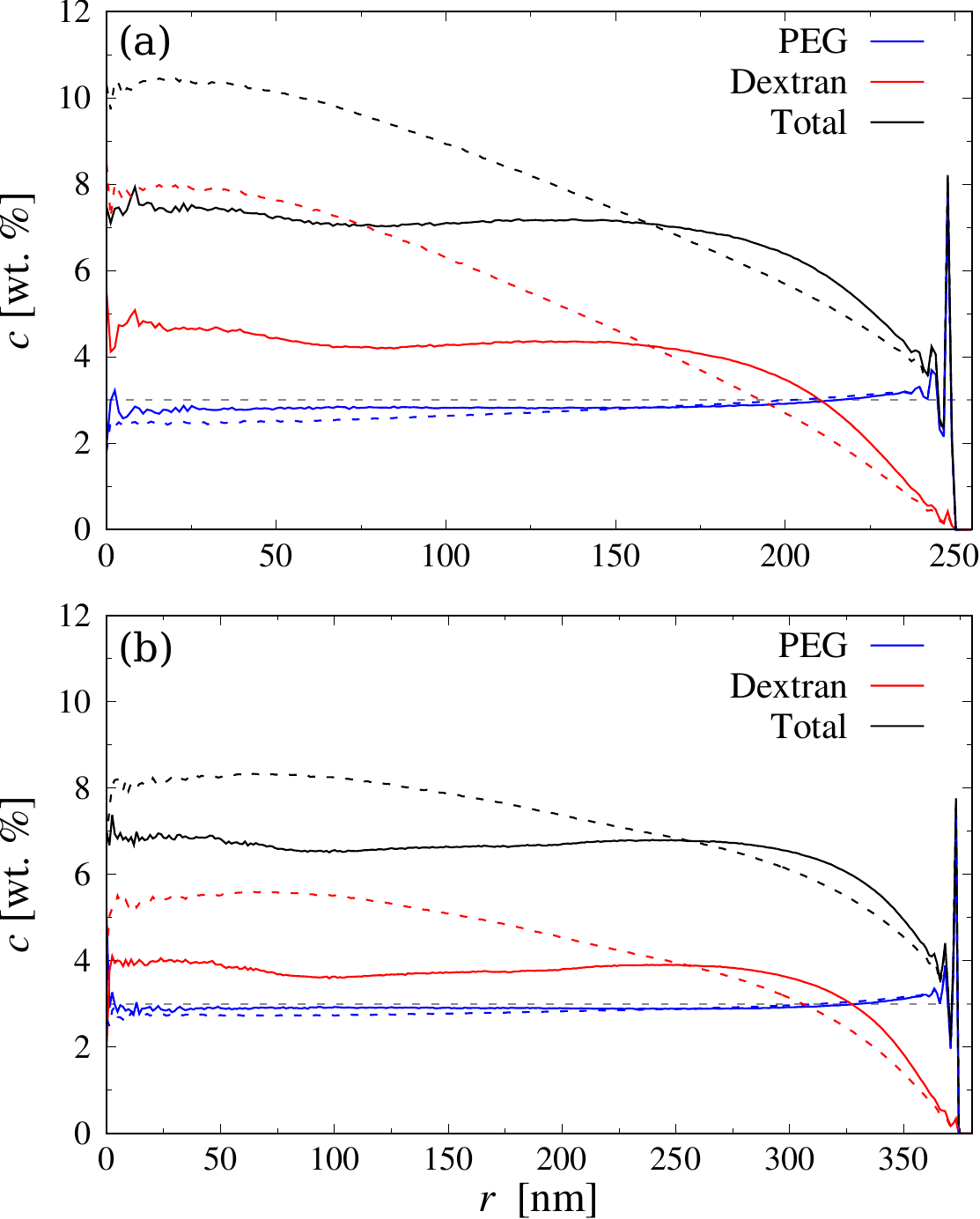}
    \caption{Radial monomer concentration profiles for droplets with (a) $R \approx 260\,{\rm nm}$ and (b) $R \approx 380\,{\rm nm}$. Solid and dashed lines show simulation results for monodisperse and polydisperse Dextran chains, respectively. The horizontal dashed grey line indicates the average polymer concentration of each species ($3\,{\rm wt}\%$).}
    \label{fig:densityProfiles}
\end{figure}

To study the spatial distribution of the PEG and Dextran polymers in the droplet, we first calculate the radial monomer concentration profiles $c(r)$ of the two species (Fig.~\ref{fig:densityProfiles}), which reveal several important features: (i) In all cases, there is a local surplus and layering of PEG near the droplet surface, which is typical for short molecules close to hard walls.\cite{snook:jcp:1978} In contrast, the long Dextran polymers are depleted from the droplet surface, because of the associated loss in conformational entropy in that region.\cite{howard:jcp:2020} In the monodisperse case, the width of this depletion zone is roughly $2\,R_{\rm g}^{\rm D} \approx 50\,{\rm nm}$, with $R_{\rm g}^{\rm D} \approx 23.8\,{\rm nm}$ being the radius of gyration of a Dextran chain at infinite dilution (see below). In contrast, the excluded region is much wider for the polydisperse case due to the broader $R_{\rm g}^{\rm D}$ spectrum. (ii) As a result, the Dextran concentration in the droplet center becomes distinctly larger than the average value ($3\,{\rm wt\%}$), reaching almost $8\,{\rm wt}\%$ for the polydisperse case in the small droplets [see Fig.~\ref{fig:densityProfiles}(a)]. By comparison, the concentration of PEG chains in the droplet center is only slightly below the average. (iii) As expected, the effect of confinement is significantly more pronounced in the smaller droplet, since the region close to the droplet surface occupies a larger volume fraction, {\it i.e.}, $1-(R-2R_{\rm g}^{\rm D})^3/R^3 \approx 0.45$ for $R=260\,{\rm nm}$ {\it vs.} $\approx 0.33$ for $R=380\,{\rm nm}$.

\begin{figure}[htb]
    \centering
    \includegraphics[width=\figwidth]{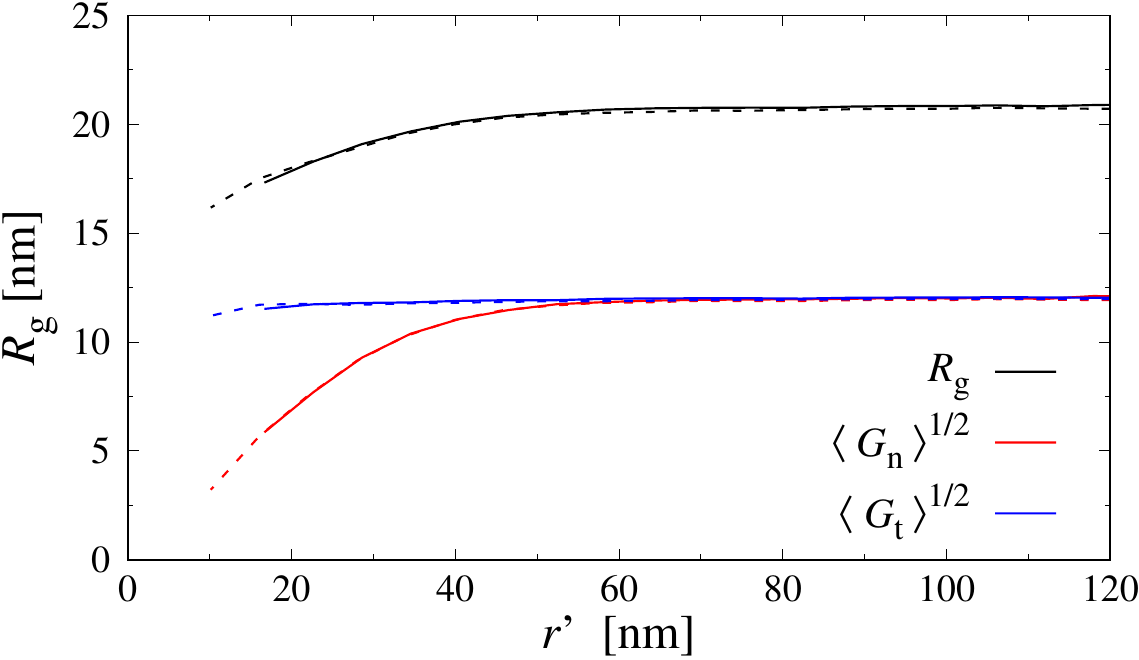}
    \caption{Radial profiles of the radius of gyration of monodisperse Dextran, $R_{\rm g}$, and its normal ($\la G_{\rm n} \ra^{1/2}$) and tangential ($\la G_{\rm t} \ra^{1/2}$) component relative to the droplet surface. Data plotted against the distance between the droplet surface and the polymer's center of mass, $r'$. Solid and dashed lines show results for $R \approx 260\,{\rm nm}$ and $R \approx 380\,{\rm nm}$, respectively.}
    \label{fig:RgProfiles}
\end{figure}

We analyze the shape of the confined Dextran chains by computing their radius of gyration tensor
\begin{equation}
    \mathbf{G} = \frac{1}{N_{\rm D}} \sum_i^{N_{\rm D}} \Delta\mathbf{r}_i \Delta\mathbf{r}_i^{\rm T},
    \label{eq:G}
\end{equation}
with $\Delta\mathbf{r}_i$ being the position of monomer $i$ relative to the polymer's center of mass. The root mean square radius of gyration is then $R_{\rm g} \equiv \la R_{\rm g}^2 \ra^{1/2} = \la G_{\rm n} + 2G_{\rm t} \ra^{1/2}$, where $G_{\rm n}$ and $G_{\rm t}$ are the components of $\mathbf{G}$ normal and tangential relative to the droplet surface, respectively. Figure~\ref{fig:RgProfiles} shows these components for monodisperse Dextran {\it vs.} the distance between the polymer's center of mass and the droplet surface. Polymers at distances larger than $\approx 2\,R_{\rm g}$ have isotropic shapes, $\la G_{\rm n} \ra^{1/2} = \la G_{\rm t} \ra^{1/2} \approx 12\,{\rm nm}$, whereas they become increasingly compressed along the normal direction as they approach the droplet surface (the tangential component is nearly constant). Further, chain conformations in the small and large droplets are almost indistinguishable.

\begin{figure*}[htb]
    \centering
    \includegraphics[width=16cm]{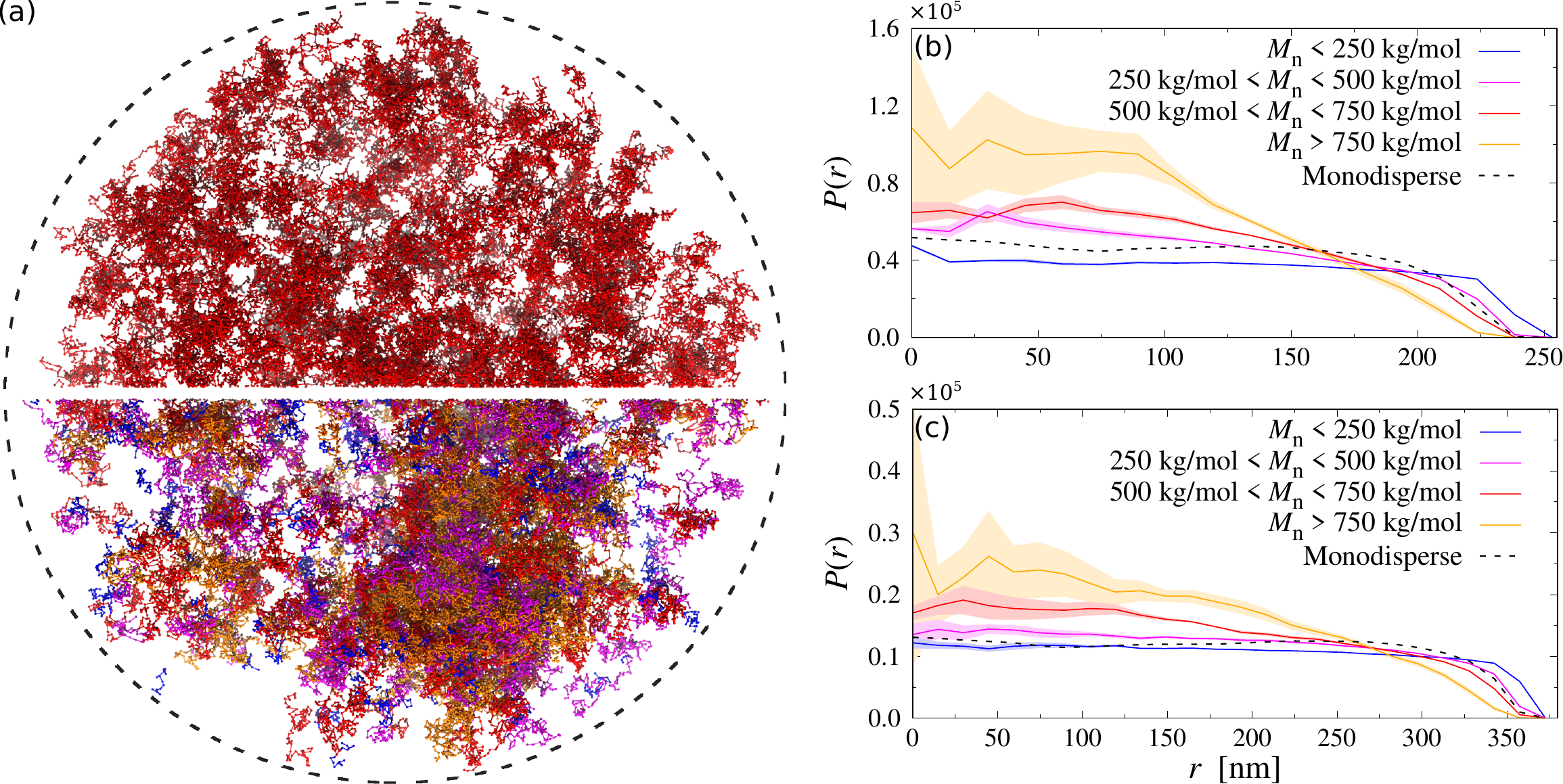}
    \caption{(a) Simulation snapshot of the confined systems ($R \approx 380\,{\rm nm}$, indicated by dashed circle). The top half shows the monodisperse case with Dextran chains colored in red. The bottom half shows the polydisperse case with Dextran chains colored according their molecular weight as in panels (b) and (c). Water and PEG particles have been omitted for clarity. (b, c) Probability $P(r)$ to find a Dextran chain in the specified molecular weight range at center-of-mass position $r$ in a droplet with (b) $R \approx 260\,{\rm nm}$ and (c) $R \approx 380\,{\rm nm}$. The shaded area indicates the standard deviation from three independent runs. The dashed black line shows the distribution of the monodisperse case.}
    \label{fig:snapSpheres}
\end{figure*}

The confinement-induced increase of the polymer concentration near the droplet center could induce a phase separation of the PEG and Dextran chains there, as the local polymer concentration might cross the binodal of the mixtures ({\it cf.} Fig.~\ref{fig:snapBulk}). Figure~\ref{fig:snapSpheres} shows simulation snapshots for the monodisperse and polydisperse systems confined in large droplets: While the monodisperse system appears to be in the mixed one-phase regime, we can clearly see large Dextran aggregates in the polydisperse case, which is in excellent agreement with recent experiments by Watanabe {\it et al.}\cite{watanabe:acsml:2022} Looking more closely at the simulation snapshot, we can see that the aggregates in the polydisperse systems primarily consist of longer Dextran chains, whereas the shorter ones still remain well dispersed. Zhao {\it et al.} observed a similar molar mass fractionation in aqueous two-phase polymer solutions of PEG and Dextran,\cite{zhao:jc:2016} where the longer Dextran chains accumulated in the Dextran-rich phase while the shorter Dextran chains were contained the PEG-rich phase.

To better understand the distinct differences between the behavior of the monodisperse and polydisperse mixtures, we determine the probability $P(r)$ of finding Dextran chains at center-of-mass position $r$, itemized by their molecular weight. These results are shown in Fig.~\ref{fig:snapSpheres}(b,c) for both droplet sizes, revealing a distinct spatial fractionation of the Dextran chains: Short polymers with $M_{\rm n} < 250\,{\rm kg/mol}$ are distributed almost homogeneously throughout the droplet, and also come much closer to the droplet surface compared to the longer chains. In contrast, longer Dextran chains are moving to the droplet center to maximize their conformational entropy. These findings are consistent with recent experiments,\cite{watanabe:acsml:2022} where Watanabe {\it et al.} inferred from surface tension measurements that short Dextran chains ($M_{\rm w} \ll 500\,{\rm kg/mol}$) preferentially adsorbed to the droplet surface. This radial partitioning of short and long Dextran chains promotes their phase separation, as longer chains have a distinctly smaller entropy of mixing compared to their shorter counterparts.\cite{rubinstein:book:2003}

To quantify the size of the Dextran condensates, we perform a cluster analysis using the density-based spatial clustering of applications with noise (DBSCAN) algorithm;\cite{dbscan} here, Dextran monomers are assigned to the same aggregate if their distance is smaller than $7\,{\rm nm}$, which roughly corresponds to the position of the first minimum of the radial distribution function $g(r)$ between Dextran particles and all other particle types. We then compute the mean aggregation number $\la M \ra = \sum_i^N MP(M)$, where $P(M)$ is the probability to find a Dextran segment in a cluster consisting of $M$ particles. To establish a baseline reference, we perform additional simulations of ideal mixtures by setting $\chi_{ij} = 0$, and determine the corresponding mean aggregation number $\la M \ra_0$. Table~\ref{tab:M} summarizes the results of all droplet simulations, which will be discussed in the following: For ideal mixtures with monodisperse Dextran chains, we find $\la M \ra_0 \approx 100$, which is comparable to the number of monomers per Dextran chain ($N_{\rm D} = 80$). This value is sensible given that $\chi_{ij} = 0$ and that the search radius of the clustering algorithm is slightly larger than the average segment length of our Dextran model ($b \approx 5.5\,{\rm nm}$). For polydisperse Dextran in the smaller droplets ($R = 260\,{\rm nm}$), the mean aggregation number increases to $\la M \ra_0 \approx 380$ due to an accumulation of longer Dextran chains in the droplet center, even at ideal conditions. This effect is considerably less pronounced in the larger droplets ({\it cf.} Fig.~\ref{fig:snapSpheres}), where we find $\la M \ra_0 \approx 110$ instead. For the non-ideal mixtures with monodisperse Dextran, we find $\la M \ra \approx 1300-1600$, which indicates the existence of small aggregates consisting of $15-20$ Dextran chains (there are, in total, 348 and 1175 Dextran chains in the small and large droplets, respectively). The mean aggregation number becomes significantly larger in polydisperse PEG-Dextran mixtures, reaching values up to $\la M \ra \approx 10000$. Interestingly, $\la M \ra$ is about two times smaller in the smaller droplets (see Table~\ref{tab:M}), which is likely a finite-size effect, as the smaller droplets contain about three times fewer Dextran chains compared to the large droplets. Nevertheless, both systems show clearly that the confinement-induced fractionation of short and long Dextran chains drives phase separation as observed in recent experiments.\cite{watanabe:acsml:2022}

\begin{table}[h]
\centering
\begin{tabular}{lcccc}
    & \multicolumn{2}{c}{Monodisperse} & \multicolumn{2}{c}{Polydisperse} \\
    \hline
$R$ & $\la M \ra$ & $\la M \ra_0$ & $\la M \ra$ & $\la M \ra_0$ \\\hline
$260\,{\rm nm}$ & $1600 \pm 60$ & $100 \pm 5$ & $5300 \pm 1200$ & $380 \pm 20$ \\
$380\,{\rm nm}$ & $1300 \pm 30$ & $97 \pm 4$ & $9800 \pm 1600$ & $110 \pm 3$ \\
\hline
\end{tabular}
\caption{Mean number $\la M \ra$ of Dextran segments in a Dextran aggregate ($\la M \ra_0$ corresponds to ideal mixtures with $\chi_{ij} = 0$).}
\label{tab:M}
\end{table}

Although we have chosen the model parameters to replicate PEG-Dextran mixtures, the rather generic nature of our coarse-grained model makes our results applicable to a wide range of different polymer mixtures. Our simulations demonstrate how the distribution of polymers is affected by confinement effects, even at good solvent conditions, with longer chains moving to the droplet center to maximize their conformational entropy. The resulting spatial inhomogeneity can drastically alter the phase behavior of the confined polymers, which is important for understanding, {\it e.g.}, the liquid-liquid phase separation of biopolymers in cellular environments. Further, our simulations provide useful guidelines for the fabrication of polymer-loaded droplets. For example, by tuning the interactions between the droplet surface and the different polymer species, one can either enhance or suppress their spatial separation, and thus control the resulting phase behavior and surface tension.

\section{Simulation Model}
\label{sec:model}
\begin{figure}[htb]
    \centering
    \includegraphics[width=6cm]{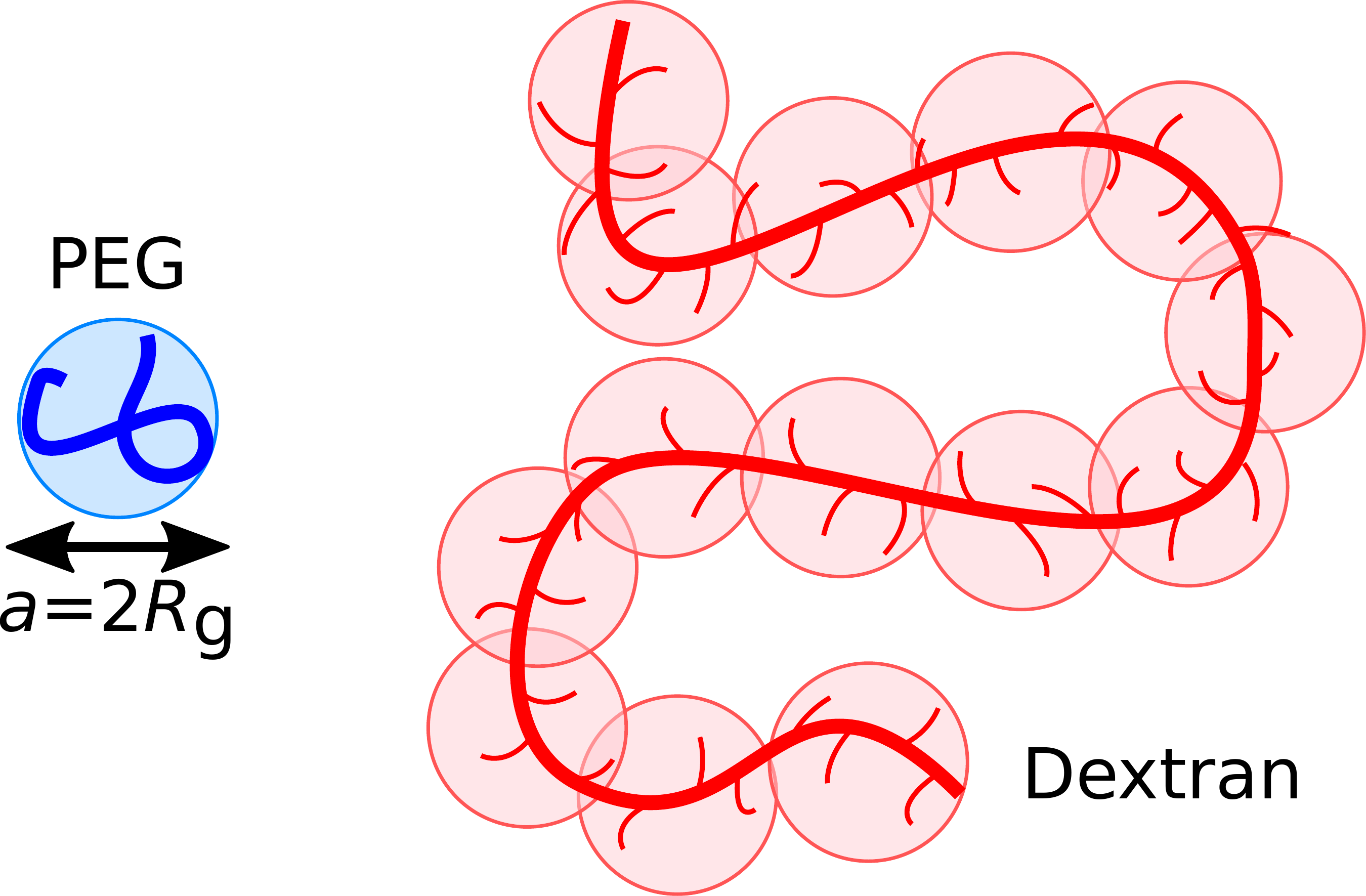}
    \caption{Schematic representation of the model, illustrating the mapping of PEG and Dextran polymers.}
    \label{fig:model}
\end{figure}

We perform dissipative particle dynamics (DPD) simulations\cite{hoogerbrugge:epl:1992, espanol:epl:1995, groot:jcp:1997} using a coarse-grained polymer model in an explicit solvent. The simulations contain three different particle types, {\it i.e.}, P (PEG), D (Dextran), and W (water), which are equal in their diameter $a$ and mass $m$. To approach experimental length- and time-scales, we model each PEG chain as a single spherical bead of diameter $a$, as shown schematically in Fig.~\ref{fig:model}. 

To establish a mapping between experiments and simulations, we first estimate the characteristic size of a PEG chain in solution: The mass and length of a Kuhn segment of PEG are $m_{\rm K}^{\rm P} = 0.137\,{\rm kg/mol}$ and $b_{\rm K}^{\rm P} = 1.1\,{\rm nm}$, respectively, resulting in $N_{\rm K} \equiv M_{\rm w}/m_{\rm K} \approx 44$ Kuhn segments for a PEG chain with molecular weight $M_{\rm w}^{\rm P} = 6\,{\rm kg/mol}$. At $\theta$-conditions, the root mean square radius of gyration can be estimated as $R_{\rm g} = b_{\rm K}(N_{\rm K}/6)^{1/2}$, resulting in a value of $R_{\rm g}^{\rm P} \approx 3.0\,{\rm nm}$. Thus, each DPD particle of diameter $a \equiv 2\,R_{\rm g}^{\rm P} = 6.0\,{\rm nm}$ represents a coil-like polymer segment. 

The Dextran chains are modeled as linear chains consisting of DPD particles with diameter $a$, because the branches of Dextran are on average typically shorter than three glucose units,\cite{larm:chr:1971} and therefore cannot be resolved at this level of coarse-graining (see Fig.~\ref{fig:model}). We determine the number of DPD beads per Dextran chain, $N_{\rm D}$, by matching $R_{\rm g}^{\rm D}$ from single chain simulations to experimental $R_{\rm g}$ measurements; in Ref.~\citenum{goerisch:jcs:2005}, $R_{\rm g}$ was derived from self-diffusion coefficient measurements of Dextran chains in water at $T=293\,{\rm K}$, with molecular weights ranging between $4\,{\rm kg/mol}$ and $464\,{\rm kg/mol}$. By extrapolating their data, we estimate $R_{\rm g} = 23.8\,{\rm nm}$ for a Dextran chain with $500\,{\rm kg/mol}$, which leads to $N_{\rm D} = 80$ for our simulation model.

Thus, each DPD particle effectively represents a coil-like polymer segment, which typically interact with each other {\it via} soft and bounded pair potentials.\cite{louis:prl:2000, likos:pr:2001, berressem:jpcm:2021} Therefore, we use the standard soft repulsion for the conservative forces acting between bonded and non-bonded DPD beads
\begin{equation}
    \mathbf{f}_{\rm m}(r) =
    \begin{cases}
    A_{ij}(1-r/a)\mathbf{\hat{r}}, & r \leq a
    \\
    0 , & r > a ,
    \end{cases}
    \label{eq:fm}
\end{equation}
where $r$ is the distance between the two particles, and $\mathbf{\hat{r}}$ is the unit vector connecting the two. The parameter $A_{ij}$ controls the repulsion strength between particles of type $i$ and $j$, and has been set according to\cite{groot:jcp:1997}
\begin{equation}
    A_{ij} = A_{ii} + 3.497\chi_{ij}
\end{equation}
with $A_{ii} = 25\,k_{\rm B}T/a$ and Flory-Huggins interaction parameter $\chi_{ij}$ (the specific values are discussed in the main text). In addition, neighboring monomers within a Dextran chain are bonded through harmonic springs with force
\begin{equation}
    \mathbf{f}_{\rm b}(r) = -k\mathbf{r} ,
    \label{eq:fb}
\end{equation}
with spring constant $k=4\,k_{\rm B}T/a^2$.\cite{groot:jcp:1998} 

In addition to these two conservative forces, all particles were subjected to pairwise dissipative and random forces
\begin{equation}
    \mathbf{f}_{\rm d}(r) = -\gamma_{ij} \omega(r) \left(\mathbf{\hat{r}} \cdot \Delta\mathbf{v}\right) \mathbf{\hat{r}} ,
    \label{eq:fd} 
\end{equation}
\begin{equation}
    \mathbf{f}_{\rm r}(r) = \sqrt{\gamma_{ij}\omega(r)} \xi \mathbf{\hat{r}} ,
    \label{eq:fr}
\end{equation}
with drag coefficient $\gamma_{ij}$, and velocity difference between two particles $\Delta\mathbf{v}$. The parameter $\xi$ is a uniformly distributed random number drawn for each particle pair, with zero mean $\la\xi(t)\ra = 0$ and variance $\la\xi(t)\xi(t')\ra = 2k_{\rm B}T\delta(t-t')$ to satisfy the fluctuation-dissipation theorem. For simplicity, we use the same drag coefficient $\gamma_{ij} = \gamma = 4.5\,\sqrt{mk_{\rm B}T}/a$ for all particles, and the standard DPD weight function\cite{groot:jcp:1997}
\begin{equation}
    \omega(r) =
    \begin{cases}
    (1-r/a)^2, & r \leq a
    \\
    0 , & r > a .
    \end{cases} 
    \label{eq:omega} 
\end{equation}

For the droplet simulations, we confine all beads to a spherical container with radius $R$ and apply a purely repulsive Weeks-Chandler-Andersen (WCA) potential\cite{weeks:jcp:1971}
\begin{align}
    U_{\rm WCA}(r') =
    \begin{cases}
    4 k_{\rm B}T \left[\left(\frac{a}{r'}\right)^{12} - \left(\frac{a}{r'}\right)^6 + \frac{1}{4}\right], & r' \leq 2^{1/6}a
    \\
    0 , & r' > 2^{1/6}a
    \end{cases},
    \label{eq:UWCA}
\end{align}
where $r'$ is the distance between the droplet surface and the center of a bead. In all simulations, the particle number density was set to $\rho = 3\,a^{-3}$. The equations of motion are integrated using a time step of $\Delta t = 0.02\,\tau$, with unit of time $\tau$. Each simulation is run for at least $10^7$ time steps, and three independent simulations are performed for each parameter set to improve statistics and determine measurement uncertainties.

\section*{Acknowledgments}
This work was supported by the Deutsche Forschungsgemeinschaft (DFG, German Research Foundation) through Project No. 470113688. M.Y. acknowledges funding by the Japan Society for the Promotion of Science (JSPS) KAKENHI (grant number 22H01188) and by the Japan Science and Technology Agency (JST) Program FOREST (grant number JPMJFR213Y). We also thank Dr. Chiho Watanabe (Hiroshima University, Japan) for fruitful discussions.

\section*{Author contributions}
\textbf{Arash Nikoubashman:} Conceptualization, Data Curation, Formal Analysis, Funding Acquisition, Investigation, Methodology, Project Administration, Resources, Software, Visualization, Writing – original draft, Writing – review \& editing.
\textbf{Miho Yanagisawa:} Conceptualization, Investigation, Project Administration, Resources, Writing – review \& editing.

\end{document}